\documentclass[twocolumn,showpacs,preprintnumbers,amsmath,amssymb, superscriptaddress]{revtex4}

\usepackage{graphicx}
\usepackage{dcolumn}
\usepackage{bm}
\usepackage[usenames]{color}    

\newcommand{\be}{\begin{equation}}
\newcommand{\ee}{\end{equation}}
\newcommand{\bea}{\vspace{0.25cm}\begin{eqnarray}}
\newcommand{\eea}{\end{eqnarray}}

\begin{document}
\title{Experimental Quantum Imaging exploiting multi-mode spatial correlation of twin beams}
\author{Giorgio Brida, Marco
Genovese, A. Meda, Ivano
Ruo Berchera\textsuperscript{*}}
\affiliation{INRIM, strada
delle Cacce 91, 10135 Torino, Italy.}
\begin{abstract}
Properties of quantum states have disclosed new and revolutionary
technologies, ranging from quantum information to quantum imaging.
This last field is addressed to overcome limits of classical imaging
by exploiting specific properties of quantum states of light. One of
the most interesting proposed scheme exploits spatial quantum
correlations between twin beams for realizing sub-shot-noise imaging
of weak absorbing objects, leading ideally to a noise-free imaging.
Here we discuss in detail the experimental realization of this
scheme, showing its capability to reach a larger signal to noise
ratio  with respect to classical imaging methods and, therefore, its
interest for future practical applications.
\end{abstract}
\pacs{42.50.Ar, 42.50.Dv, 42.50.Lc,03.65.Wj} \maketitle
\section{Introduction}
The possibility of manipulating quantum states as atoms, photons,
etc. has recently fostered the developing of quantum technologies
\cite{1,2,3,4,5,6,7,8,9,10,11,12,13,14,Kol2007,BrambPRA2004,15,15bis},
with very promising opportunities for future widespread
applications. Among quantum technologies the ones based on photons,
in particularly exploiting entanglement as a resource, have produced
the most impressive results, such as quantum cryptography
\cite{2,3}, teleportation \cite{4,5,6}, linear-optics quantum
computation \cite{7}, entanglement distillation \cite{8} and
purification \cite{9,10,11}, quantum metrology \cite{15}. The use of
quantum properties of light for overcoming limits of classical
imaging is one of the most promising \cite{12,13,14}. In particular
it was suggested that, by exploiting the strong correlation in noise
of entangled modes of light produced by Parametric Down Conversion
(PDC), the image of a weak absorbing object in one branch,
eventually previously hidden in the noise, can be restored by
subtracting the spatial noise pattern measured in the other branch,
a technique that has been christened Sub Shot Noise Quantum Imaging
(SSNQI) \cite{19}. When operating under shot noise regime (avoiding
the limits of the need of background subtraction \cite{20} or single
spatial mode operation \cite{16,17,18,masha}) the method allows in
principle a full reconstruction of the absorption pattern of an
object with a sensitivity superior to that available with classical
techniques at the same illumination level. In order to reach this
goal in view of important practical applications one should reach a
very high level of spatial quantum correlation. In \cite{19} we have
demonstrated that this level can be effectively reached and, then,
in \cite{nphot} we have presented a proof of principle of the sub
shot noise quantum imaging scheme. Here, in order to provide all the
necessary information, we present further data, discussing the
details of the experimental realization. In particular, we show how
to reach a degree of spatial correlation needed for SSNQI
performances better than the corresponding classical scheme, i.e. a
differential classical imaging (DCI) scheme in which, instead of the
twin beams of PDC, two classically correlated beam are used. This
achievement demonstrates that an experimental setup for SSNQ imaging
is feasible and the practical application of the method can be
upcoming. In Sec.\ref{The Scheme} we provide a basic introduction to
the multi-mode spatial squeezing and the idea of SSNQI exploiting
PDC, while Sec.\ref{Experimental setup and sub-shot noise regime}
describes the experimental set-up and the spatial sub-shot-noise
measurements. Sec.\ref{Experimental imperfection and their
correction} considers some effects that spoil the spatial
sub-shot-noise regime and their possible corrections. Finally, in
Sec.\ref{quantum vs classical imaging} we present our quantum
imaging experiment, and two different figures of merit that compare
the performances of SSNQI and DCI. Some conclusion and remarks can
be found in Sec.\ref{Conclusion}

\section{The Scheme}\label{The Scheme}

The scheme \cite{14} for achieving quantum imaging of weak absorbing
objects, is based on considering two spatially correlated areas of
Parametric Down Conversion (PDC) light. This is a non-linear optical
phenomenon \cite{1,15bis} where a pump beam photon "decays" in two
lower frequencies photons, usually called signal (s) and idler (i),
strongly correlated in frequency and momentum since they have to
conserve the energy and momentum of the original pump photons. In
particular, the conservation of the transverse (to the propagation
direction) component of the momentum traduces in a correlation
between the photon number $N_{i} (\mathbf{x})$ and
$N_{s}(-\mathbf{x})$ ideally detected in any pairs of symmetrical
positions $\mathbf{x}$  and $-\mathbf{x}$ in the far field plane
(note: we restrict to wavelength near degeneracy wavelength
$\lambda_{s}=\lambda_{i}$). The degree of correlation is quantified
by the Noise Reduction Factor (NRF)
\begin{eqnarray}\label{sigma}
\sigma&\equiv&\frac{\left\langle\delta
^{2}(\widehat{N}_{i}-\widehat{N}_{s})\right\rangle} {\left\langle
\widehat{N}_{i}+\widehat{N}_{s}\right\rangle}\\\nonumber
&=&\frac{\langle \delta^{2} \widehat{N}_s\rangle+\langle
\delta^{2} \widehat{N}_i\rangle-2\langle \delta \widehat{N}_s;\delta
\widehat{N}_i\rangle} {\left\langle
\widehat{N}_{s}+\widehat{N}_{i}\right\rangle},
\end{eqnarray}
where $\langle\delta^{2}\widehat{O}\rangle=\langle
\widehat{O}^{2}\rangle-\langle\widehat{O}\rangle^{2}$ is the mean
square fluctuation of a generic operator $\widehat{O}$. The
normalization in Eq. (\ref{sigma}) represents the Shot Noise Limit
(SNL) $\langle \widehat{N}_{s}+\widehat{N}_{i}\rangle$, defined as
the level of fluctuation associated to the difference between
coherent beams. In terms of the transmission $\eta$ of the optical
channel (including the quantum efficiency of the detector) the
theoretical value of NRF for PDC is $\sigma=1-\eta$ (considering for
simplicity balanced  losses $\eta_{s}=\eta_{i}=\eta$ that leads to
symmetrical statistical properties of the detected beams).
Therefore, in an ideal case in which $\eta\rightarrow 1$, $\sigma$
approaches zero. On the other side, for the subtraction of two
classical beams the degree of correlation is bounded by
$\sigma_{class}\geq1$, where the lowest limit $\sigma_{class}=1$ is
reached for coherent beams, or correlated beams, generated by a
single classical beam separated by a balanced beam splitter. In this
case the subtraction of the detected intensities allows to eliminate
the classical excess noise contained in the source, while the shot
noise survives, leading to $\langle\delta
^{2}(\widehat{N}_{i}-\widehat{N}_{s})\rangle=2\langle\widehat{N}_{s}\rangle$.
The idea of sub-shot noise quantum imaging (SSNQI), depicted in Fig.
\ref{fig1}, consists in measuring the intensity pattern on one
branch (e.g. the signal one), where the object has been inserted,
and then subtracting the correlated noise pattern measured in the
other branch (the idler one) that does not interact with the object.
The number of photon detected in presence of the object in the
signal region is $\langle
N'_{s}(\mathbf{x})\rangle=[1-\alpha(\mathbf{x})]\langle
N_{s}\rangle$ where $\alpha(\mathbf{x})$ is the absorption in the
position $\mathbf{x}$.

Therefore, in the SSNQI scheme we evaluate the absorbtion as
\begin{equation}\label{alfa}
\alpha(\mathbf{x})=\langle[N_{i}(-\mathbf{x})-N'_{s}(\mathbf{x})]\rangle/\langle N_{i}\rangle
\end{equation}
where implicitly we always consider $\langle N_{i}\rangle=\langle N_{s}\rangle$.

The Signal to Noise Ratio (SNR) is improved because the spatial
noise that affects the single beam is washed away by the
subtraction. In Ref.\cite{14}  the theoretical capability of the
described method exploiting PDC has been compared with the ones of
the corresponding differential classical scheme in which a coherent
beam is split by a 50\% beam splitter. It was shown that this leads
to a ratio between the SNR in quantum and in the differential
classical imaging of

\begin{equation}\label{Rdcl}
R^{DCI} =
\frac{SNR_{SSNQI}}{SNR_{DCI}}=\sqrt{\frac{(2-\alpha)}{\alpha^{2}E_n+
2\sigma(1-\alpha) + \alpha}}
\end{equation}
for the same number of photons. We have introduced the excess noise
$E_n= (\langle \delta^{2}N_{i}\rangle-\langle N_{i}\rangle )/
\langle N_{i}\rangle$ of the PDC, i.e. the noise that exceeds the
SNL, also referred to as standard quantum limit. For multi-thermal
statistics of the single branch of PDC light, the excess noise is
roughly given by $\langle N\rangle /\mathcal{M}$, with $\mathcal{M}$
the number of modes. Equation \ref{Rdcl} shows that, when the excess
noise is negligible (in particular $\alpha^{2}E_n\ll1$), the SSNQI
presents an advantage respect to a classical differential imaging
for a weak absorbing object ($\alpha\rightarrow0$) as soon as
$\sigma<1$. This formula can be generalized to the case of
multi-thermal super Poissonian source for the differential classical
scheme (instead of only considering coherent light) just including
in the numerator under square root the term $E^{class}_n
\alpha^{2}$, that represents the contribution due to the excess
noise $E^{class}_n$ of the classical light.

Furthermore, it can be demonstrated that for $\sigma<0.5$ the
quantum imaging technique is more advantageous than an ideal direct
classical scheme, and therefore it can be a useful resource for
practical applications \cite{iv}.

\section{The experimental realization of the Sub-Shot-Noise}\label{Experimental setup and sub-shot noise regime}

\begin{figure}[tbp]
\includegraphics[width=0.5\textwidth]{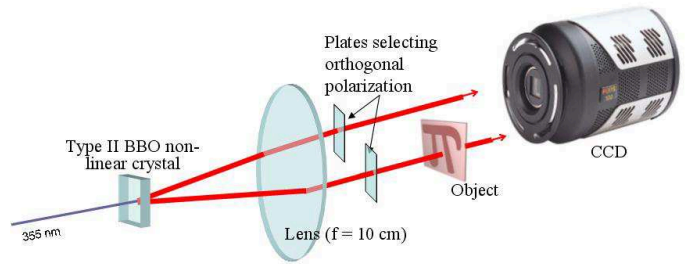} \caption{Experimental set up. A 355 nm laser beam
pumps a type II BBO crystal producing PDC. After eliminating UV
beam, one correlated beam cross a weak absorbing object  and it is
addressed to  a CCD array. The other beam (reference)  is directly
addressed to another area of CCD camera. The total transmittance of
the optical path is evaluated to 70\%. }\label{fig1}
\end{figure}

In our experimental setup a type II BBO non-linear crystal ($l=7$
mm) is pumped by the third harmonic (355 nm) of a Q-switched Nd:Yag
laser. The pulses have a duration of $T_{pump}=5$ ns with a
repetition rate of 10 Hz and a maximum energy, at the selected
wavelength, of about 200 mJ. This choice for the pulse duration is
motivated by the indications of Ref. \cite{14} for realizing a
set-up suited for a test of SSNQI, overcoming the limits of Ref.
\cite{20}. After eliminating with a prism the residual of first and
second harmonic components, the pump beam crosses a spatial filter
(a lens with a focal length of 50 cm and a diamond pin-hole, 250
$\mu$m of diameter), in order to eliminate the non-gaussian
components and then it is collimated by a system of lenses to a
diameter of $w_{p}=1.25$ mm. After the crystal, the pump is stopped
by a couple of UV mirrors, transparent to the visible (T=98\% at 710
nm), and by a low frequency-pass filter (T=95\% at 710 nm). The down
converted beams (signal and idler) are separated in polarization by
two polarizers (T=97\%) and finally the far field is imaged by a CCD
camera \ref{fig1}. We used a $1340X400$ CCD array, Princeton
Pixis:400BR (pixel size of 20 $\mu$m), with high quantum efficiency
(around 80\%) and low noise in the read out process ($\Delta=4$
electrons/pixel). The CCD exposure time is set by a mechanical
shutter to 90 ms, thus each image cached in this time window
corresponds to the PDC emission generated by a single shot of the
laser. The far field is observed at the focal plane of the lens
($f=10$ cm) in a $f-f$ optical configuration, ensuring that we image
the Fourier transform of the crystal exit surface. Therefore, a
single transverse mode $\mathbf{q}$ of the down converted photons is
associated to a single point $\mathbf{x}=(\lambda f/2\pi)\mathbf{q}$
of the image.

We expect a strict correlation in the number of photons
$\widehat{N}_{i}$ and $\widehat{N}_{s}$ detected at any couple of
symmetric positions with respect to the center of symmetry,
providing the detection areas are larger than the typical coherence
area \cite{19}. The coherence area $\mathcal{A}_{coh}$, i.e. the
uncertainty on the relative position of the correlated photons in
the far field, depends on the pump angular bandwidth \cite{14}, and
on the parametric gain \cite{21,22}. In our framework the detectors
are the pixels of the CCD array and therefore the single pixel
should be of the same order of magnitude of the coherence area or
bigger. By contrary, if the pixel is smaller, the photons detected
at the pixel in the position $\mathbf{x}$ will have the correlated
ones spread on several pixels around the position $-\mathbf{x}$.
This spoils the pixel-pixel correlation.

For this reason, and in order to reduce the contribution of the read
noise of the CCD, it is convenient to perform an hardware binning of
the physical pixels. It consists in grouping the physical pixels in
squared blocks NxN, each of them being processed by the CCD
electronics as single "superpixel" (the photons collected by the
superpixel are the sum of the photons of each pixel, whereas the
read noise is just slightly increased with respect the one of the
single pixel). As described in \cite{19}, following the previous
argumentation, the size of the superpixel $\mathcal{A}_{pix}$ must
be at least of the order of the coherence area, although the ideal
condition is $\mathcal{A}_{pix}/\mathcal{A}_{coh}\gg1$. In the
present paper we consider three different superpixel sizes: 12x12,
24x24 and 32x32, being the measured coherence area
$\mathcal{A}_{coh}\sim(6\:\mathrm{pixel})^{2}=(120\mu\mathrm{m})^{2}$.

\begin{figure}[tbp]
\includegraphics[width=0.5\textwidth]{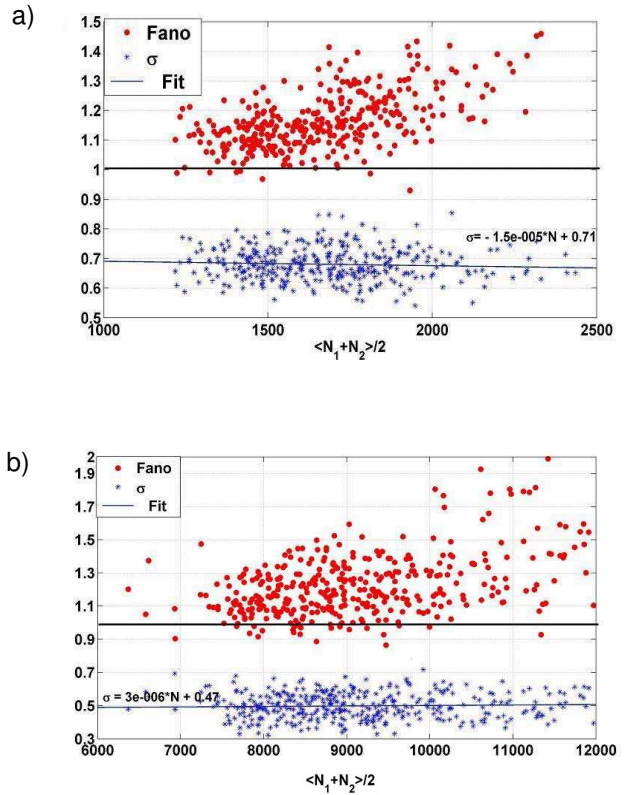}
\caption{Degree of Correlation and Fano factor versus the number of
photons over a sample of 400 frames analyzed. Fig2(a) presents the
data for binning 12x12, i.e. a superpixel with size  $(240\mu
m)^{2}$ and regions containing 320 superpixels. The mean value of
correlation degree for this range of photon number is
$\sigma=0.681\pm0.003$ and the standard deviation of the sample is
0.06. The mean value of the Fano factor is $F=1.17$ and the standard
deviation of the population is 0.13. Fig2(b) presents the same
analysis for binning 24x24 (superpixel size $(480\mu m)^{2}$) and
regions containing 56 superpixel. The mean value of correlation
degree is $\sigma=0.498\pm0.004$ and the standard deviation 0.08.
Fano factor has mean value $F=1.26$ and standard deviation of
0.35.}\label{fig2}
\end{figure}

The number of temporal modes collected in one shot image is
$\mathcal{M}_{temp}=(T_{pump}/T_{coh})\sim5\cdot10^{3}$, estimating
the coherence time $T_{coh}$ of the PDC process around one
picosecond \cite{1}. The number of spatial modes collected by the
superpixel is
$\mathcal{M}_{spatial}\sim\mathcal{A}_{pix}/\mathcal{A}_{coh}$
depends only on its size, since $\mathcal{A}_{coh}$ is fixed by the
pump transverse size. From the total number of modes $
\mathcal{M}=\mathcal{M}_{temp}\otimes
\mathcal{M}_{spatial}\sim10^{4}\div10^{5}$ and the mean photon
number per pixel one expect the excess noise of the multi-thermal
statistics $E_n\equiv\left\langle
N_{s}\right\rangle/\mathcal{M}\sim0.1$. It can be seen as the number
of photons per mode, and here is fixed, since we keep fixed (aside
unwanted fluctuation pulse-to-pulse) the power of the laser.
Therefore the multi-thermal statistics of the source in the single
channel is quasi Poissonian, and this is confirmed by the Fano
factors $F=1+E_{n}=\langle \delta^{2}N_{i}\rangle/\langle
N_{i}\rangle $ presented in Fig.\ref{fig2}.

The first step of the measurement consists in evaluating the degree
of correlation between pixel pairs belonging to large areas of PDC
emission, without object. Chosen two correlated regions of PDC in
the far field $A_{s}$ and $A_{i}$ of about $(3mm)^2$ size, we
evaluate the degree of correlation in Eq.(\ref{sigma}) by performing
spatial averages over the ensemble of correlated pixels belonging to
the two regions. The measurement procedure is described in
\cite{19}. Fig.\ref{fig2} shows the results obtained for a sample of
400 images (each one corresponding to a pump shot and to a point in
the graphs) for the two binning size. In both cases the degree of
correlation is well below the SNL. As expected, the result for
binning 24x24  is better than for 12x12, and the last is better then
the one reported in \cite{19}, anyway. Thus, we are in the regime
$\sigma<1$ needed for appreciating an improvement of the SSNQI with
respect to the classical differential scheme.

\section{Experimental imperfection and their correction \label{Experimental imperfection and their correction}}
In this section we aim to illustrate the experimental challenges for
achieving a strong local sub-shot-noise correlation with a large
number of detected photons over a large number of pixels at the same
time. Actually working with high photon number is demanded for
reducing the effect of the background. In fact the background
(electronic noise of the CCD and straylight) can be corrected a
posteriori on average, but the single image remains affected by this
noise. Therefore we need to work in a condition of
$\langle\delta^{2} B\rangle/\langle
\widehat{N}_{i}+\widehat{N}_{s}\rangle\ll1$, where we indicate as
$B$ the number of background photons. The main sources of
experimental excess noise are the following:

1) the intensity gradients in the regions $A_{s/i}$ of interest,

2) the error in the determination of the center of symmetry of the spatial correlation,

3) the background noise due to the straylight and the read-out noise of the CCD.

We analyze them in detail in the following subsections.

\subsection{Intensity gradients}
Up to now, we have considered an ideal situation in which the
quantum efficiency $\eta$ is constant and equal in the two optical
paths.

From an experimental point of view, we had to cope with several
sources of spatial dishomogeneity that contribute to the
experimental excess noise and that can hide the quantum correlation,
especially when large areas of PDC emission around degeneracy
(710nm) are considered.

In the angular bandwidth corresponding to an area of (3 mm)$^2$
close to collinear direction, lies a frequency bandwidth of about
10-20 nm. The response of the filters and the CCD quantum efficiency
have a frequency dependence even within this small range. This
generates a spatial gradient in the intensity pattern. For instance
the CCD sensitivity, that has a maximum around 800 nm, varies of
about 3\% in our range of wavelengths. The bandpass filter (see
Fig.\ref{fig1}) has a maximum of transmittance of 95\% at 710 nm,
decreasing of few parts percent at the border of our frequency
region. These effects are even more complicated by the non-linear
dependence of the frequency on the emission angle in the
phase-matching function. Even the optical transmission at different
angles can be influenced by the polarization. Moreover, the spatial
statistics of the detected light is affected by spatial
imperfections of the optics, by the presence of dust and by the
inhomogeneity of efficiency of the CCD pixels (a minor
contribution). All these effects contribute to an increasing of the
excess noise and, consequently, a deterioration of the noise
reduction factor.

We modeled it by introducing a spatial dependence
of the quantum efficiency $\eta_{j,k}$, with j=s,i and $k$
identifying the symmetrical pixels in the position $\mathbf{x}_{k}$ and
$-\mathbf{x}_{k}$ of the regions $A_s$ and $A_i$ respectively. The
expectation values of the number of photons on the pixel $k$ is:
\begin{equation}\label{eN}
\langle \hat{N}_{j,k}\rangle= \eta_{j,k}\langle \hat{n}\rangle
\end{equation}
with $\langle \hat{n}\rangle$ the number of photon impinging a pixel
in each frame. The associated variance and covariance are
\cite{Commento}
\begin{eqnarray}
\label{s_eN}
\langle \delta^{2} \hat{N}_{j,k}\rangle&=&
\eta_{j,k}\langle \hat{n}\rangle+\eta_{j,k}^{2 }\langle
\hat{n}\rangle^{2}/\mathcal{M}\\
\label{ss_eN} \langle \delta \hat{N}_{s,k}\delta
\hat{N}_{i,k}\rangle&=& \eta_{s,k}\eta_{i,k}(\langle \hat{n}\rangle+
\langle \hat{n}\rangle^{2}/\mathcal{M}).
\end{eqnarray}
Using Eq. (\ref{eN}), Eq. (\ref{s_eN}) and  Eq. (\ref{ss_eN}) in Eq. (\ref{sigma}) we
can calculate the theoretical value of the NRF:
\begin{equation}\label{sigma_eff_th}
\sigma_k =
1-\eta_{+,k}+\frac{1}{2}\frac{\eta^{2}_{-,k}}{\eta_{+,k}}\left(\frac{\langle
n \rangle}{\mathcal{M}}+\frac{1}{2}\right)
\end{equation}
where $\eta_{+,k}= \frac{\eta_{s,k}+\eta_{i,k}}{2}$ is the mean
value of the quantum efficiencies of signal and idler arm and
$\eta_{-,k}= \eta_{s,k}-\eta_{i,k}$ is their difference. It is worth
to observe that the degree of correlation is deteriorated by the
unbalancing in the quantum efficiencies $\eta_{-,k}$ on the two
paths. Due to the high pulse-to-pulse instability of the Q-switch
laser, in our experiment it is convenient to estimate the NRF in the
single frame obtained by a single laser shot. A single frame
contains several pairwise spatial modes of the radiation, since the
PDC is spatially broadband \cite{BrambPRA2004}; therefore, pixels
belonging to regions $A_i$ and $A_s$ are pairwise correlated in the
number of photons, and the pairs can be considered as independent if
the condition $\mathcal{A}_{pix}\gg \mathcal{A}_{coh}$ is fulfilled.
We estimate NRF as defined in Eq.(\ref{sigma}), where the quantum
mean values are evaluated by averages over the ensemble of the pixel
pairs. The operator associated to the estimation of the mean photon
number is
\begin{equation}\label{S}
\widehat{\mathcal{N}}_j=\frac{1}{\mathcal{R}}\sum_{k=1}^{\mathcal{R}}
\widehat{N}_{j,k}
\end{equation}
where $\mathcal{R}$ is the number of pixels in the region. According
to Eq. (\ref{eN}), the expected value of $\hat{\mathcal{N}}$ becomes
\begin{equation}\label{eS}
\langle\hat{\mathcal{N}}_j\rangle=\frac{1}{\mathcal{R}}
\sum_{k=1}^{\mathcal{R}}
\eta_{j,k}\langle\hat{n}\rangle=\overline{\eta}_{j}\langle
\hat{n}\rangle.
\end{equation}
 Similarly, the operator for the
variance of the photon number is defined as:

\begin{equation}\label{dS}
\widehat{\mathcal{S}}^{2}_j=\frac{1}{\mathcal{R}}\sum_{k=1}^{\mathcal{R}}
(\widehat{N}_{j,k}-\langle\hat{\mathcal{N}}_j\rangle)^{2}
\end{equation}
Using Eq. (\ref{s_eN}), its expected value is  $\langle
\widehat{\mathcal{S}}_j^{2}\rangle= \overline{\eta}_{j}\langle
\hat{n}\rangle+\left(\overline{\eta}_{j}^{2}+V(\eta_{j})/\mathcal{M}\right)
\langle \hat{n}\rangle^{2}/\mathcal{M}+V(\eta_{j})\langle
\hat{n}\rangle^2$ (here we exclude correlation between the
efficiency distribution in the two regions). We observe that
performing the spatial statistics in order to estimate the
theoretical quantities introduces a term of excess noise depending
on the spatial variance $V(\eta_{j})$ of the detection efficiency in
the region $A_j$. At the same time, we can define the operator
associated to the estimation of covariances as
\begin{equation}\label{ddS}
\widehat{\mathcal{S}}^{2}_{s,i}=
\frac{1}{\mathcal{R}}\sum_{k=1}^{\mathcal{R}} (
\hat{N}_{s,k}-\langle \hat{\mathcal{N}}_{s}\rangle)(
\hat{N}_{i,k}-\langle \hat{\mathcal{N}}_{i}\rangle)
\end{equation}
According to (\ref{ss_eN}) and to (\ref{eS}),its expected value is
$\langle\widehat{\mathcal{S}}^{2}_{s,i}\rangle=\overline{\eta}_{s}\overline{\eta}_{i}\left(\langle
\hat{n}\rangle+ \frac{\langle
\hat{n}\rangle^2}{\mathcal{M}}\right)$. Using the expectation values
defined above, we can now estimate the experimental noise reduction
factor:
\begin{eqnarray}\label{sigma_eff} \sigma^{exp} =
1-\overline{\eta}_{+}+\frac{1}{2}\frac{\overline{\eta}^2_{-}}{\overline{\eta}_{+}}\left(\frac{\langle
n
\rangle}{\mathcal{M}}+\frac{1}{2}\right)+\nonumber\\+\left[\frac{V(\eta_s)+V(\eta_i)}{2\overline{\eta}_{+}}\right]\left(\langle
n \rangle+\frac{\langle n \rangle}{\mathcal{M}}\right)
\end{eqnarray}

Comparing the expected value of the NRF $\sigma^{exp}$ estimated in
the single frame with the theoretical value (\ref{sigma_eff_th})
that refers to the single pixel pair, a new term of excess noise
depending on the spatial variance of the efficiency appears.
Although the variance is usually rather small, this term can be
relevant (even the dominant one) when the number of photons per
pixel becomes as high as we deal in our working condition, and needs
to be addressed.

This additional term can be corrected a posteriori by compensating
the channel losses represented by $\eta_{j,k}$. Focusing, as usual,
on the two regions of interests, the compensation consists in
multiplying each matrix-image by a sort of "flat field" matrix
$g_{j,k}$. This is obtained by the sum of $\mathcal{Q}$ single shot
images $F_{j,k}=\sum_{q=1}^{\mathcal{Q}} N^{(q)}_{j,k}$ normalized
for the mean value in the region $j$ as

\begin{equation}\label{flat field}
g_{j,k}^{-1}=\frac{F_{j,k}}{\frac{1}{\mathcal{R}}\sum_{k=1}^{\mathcal{R}}
F_{j,k}}=\frac{\eta_{j,k} }{\overline{\eta}_j}.
\end{equation}

The flat field correction corresponds, in the theoretical model, to the substitution $N_{j,k}\rightarrow
g_{j,k}N_{j,k}$ in the expressions (\ref{S}), (\ref{dS}) and (\ref{ddS}).
Consequently, applying the flat field compensation as in Eq. (\ref{flat field}), the experimental NRF becomes:
\begin{equation}\label{sigma_eff_FF}
\sigma^{F} =
1-\overline{\eta}_{+}+\frac{1}{2}\frac{\overline{\eta}^{2}_{-}}{\overline{\eta}_{+}}\frac{\langle
n \rangle}{\mathcal{M}}
\end{equation}
that is a good estimation of the average NRF (\ref{sigma_eff_th})
over the two large regions. Thus, the application of this
compensation for the losses disuniformity allows in principle to
retrieve the expected quantum mean value of the NRF. Fig.
\ref{plot1} and Fig. \ref{plot2} report the NRF and the Fano factor
in function of binning size and their values corrected by the flat
field and for background noise. From the figures we notice that the
effect of disuniformity, $V(\eta_{j})$, dominates and the flat field
correction becomes fundamental. In fact, an increasing of the
superpixel size corresponds to an increment of the number of photons
and, consequently, of the excess noise generated by $V(\eta_{j})$ in
Eq.(\ref{sigma_eff}). It is important to stress that this
compensation is the same for all the images, since the
transmission-detection efficiency distribution is always the same.
Therefore, this technique can be applied successfully even in the
imaging scheme, when an object is inserted.

\subsection{Determination of the center of symmetry and sizing of the superpixel}

We observe that the theoretical prediction $\sigma=1-\eta$ relies on
the assumption that each pixel of $\mathcal{A}_{s}$ detects all the
spatial modes correlated to the ones collected by the corresponding
pixel in $\mathcal{A}_{i}$, and viceversa. Otherwise, the presence
of uncorrelated modes in the two regions would not provide a
complete cancelation of the excess noise by the subtraction
\cite{masha}. In particular it can be demonstrated that
\begin{equation}\label{uncorr-modes}
\sigma'=1-\eta\frac{\mathcal{M}_{c}}{\mathcal{M}_{c}+
\mathcal{M}_{u}}\left(1-\frac{\mathcal{M}_{u}}{\mathcal{M}_{c}}E_{n}\right).
\end{equation}
with $\mathcal{M}_{c/u}$ the number of spatial
correlated/uncorrellated modes detected by a pair of symmetric
pixels. When the center of symmetry (CS) is estimated exactly,
$\mathcal{M}_{u}=0$ and one retrieves the theoretical lower bound
$\sigma'=\sigma$. Therefore, experimental precise positioning of the
regions and accurate determination of the CS is fundamental. In
\cite{23} it is demonstrated by realistic simulation that an high
level of correlation is obtained by finding the CS with sub-pixel
resolution. In the present work the CS used in a measurement with a
certain NxN binning grid (N=12 or N=24), is evaluated previously by
using a more resolved grid (typically 2x2) as follows. Considering
the region $A_s$ we find the proper position of the correlated
region $A_i$ evaluating the noise reduction factor $\sigma$ as a
function of the position of $A_i$ (moved by integer steps). The
optimal position corresponds to a dip in the value of $\sigma$.
Finally, taking two correlated pixels $x_s$ and $x_i$, the center of
symmetry is $x_c=\frac{|x_i-x_s|}{2}$. A different approach is based
on a procedure of NRF optimization by micro-positioning of the CCD
array \cite{BDGRR2010in prep}.

However, for a given error in the determination of the CS,
increasing $\mathcal{A}_{pix}$ reduces the ratio between the
uncorrelated and the correlated modes. Therefore, an appropriate
determination of the minimum size of the superpixel can reduce this
bias under an accepted threshold. Fig. \ref{plot1} shows clearly a
strong dependence of the NRF (after flat field correction) from the
pixel size. This behaviour is well explained by a geometrical model
based on Eq. (\ref{uncorr-modes}).

\subsection{Background noise}
\begin{figure}
\includegraphics[width=0.5\textwidth]{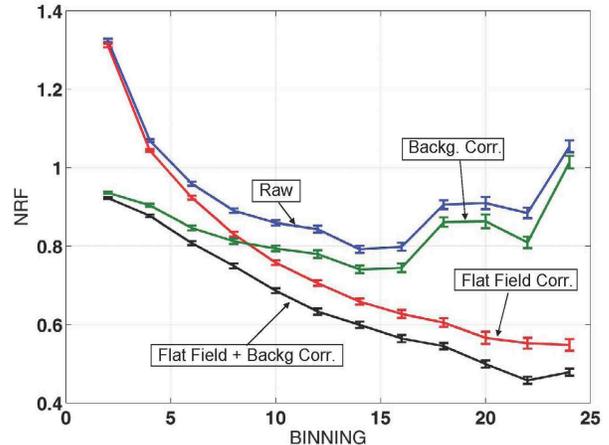}
 \caption{The plot reports the value of the
Noise Reduction Factor as a function of binning size N. The red
(continuous) line refers to raw data, without any noise correction.
The corresponding data after the background subtraction are reported
in the green (dashed)line. The red (dotted) line refers to data
corrected with Flat Field and the black (dotted dashed)  line data
takes into account both Flat Field and background
corrections.}\label{plot1}
\end{figure}

\begin{figure}
\includegraphics[width=0.5\textwidth]{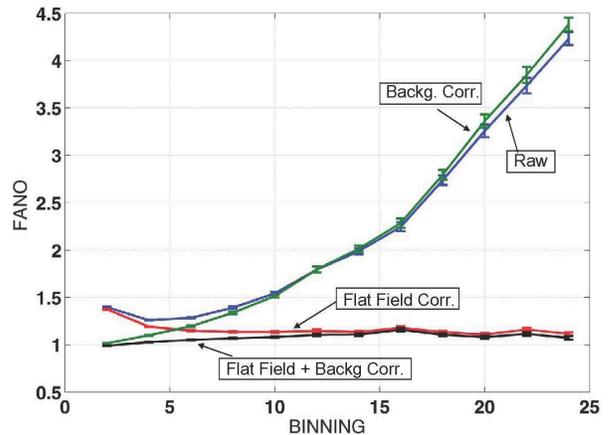} \caption{The plot reports the value of the
Fano factors as a function of the binning size N. The colors are
according to Fig. \ref{plot1} }\label{plot2}
\end{figure}

Another source of deterioration of the spatial correlations is the
background noise. The two main sources of background noise are the
electronic noise in the read-out and digitalization process and the
straylight. The electronic noise of the CCD is $\Delta=4$
photoelectron/pixel. This contribution can be strongly reduced in
terms of signal to noise ratio when binning is performed. Straylight
light is mainly due to the broadband fluorescence of the crystal,
mirrors and filters hit by the strong UV pulse of the laser and to
the residual UV light itself. The background noise can be estimated
independently by acquiring an image without PDC light, simply by
rotating the pump polarization of 90 degrees. The experimental value
of the NRF (\ref{sigma}) can be corrected a posteriori  taking into
account the mean value of the background:

\begin{equation}\label{sigma cor}
 \sigma^{B}=\frac{\left\langle\delta
^{2}(\widehat{N}_{i}-\widehat{N}_{s})\right\rangle-\left\langle\delta^{2}
B_{s}\right\rangle-\left\langle\delta^{2} B_{i}\right\rangle}{\left\langle
\widehat{N}_{i}+\widehat{N}_{s}\right\rangle-\left\langle B_{s}+B_{i}\right\rangle}
\end{equation}

where $\langle B_{j}\rangle$ is the mean photon numbers of the area
$A_{j}$ of the background image and $\langle \delta^{2}
B_{j}\rangle$ is its variance. The same correction can be applied to
the Fano factors. From Fig. \ref{plot1} we notice that the
electronic noise is the main source of deterioration of sigma for
low binning size (small $\mathcal{A}_{pix}$), while for larger
binning the effect of disuniformity, $V(\eta_{j})$, dominates. This
can be explained considering that the read-out noise of the pixel
(or superpixel) is a fixed value, independent on the binning.
Nonetheless larger superpixels detect more photons, reducing the
effect of read-out noise ($\langle\delta^{2}B_{j}\rangle/\langle
\widehat{N}_{j}\rangle\ll1$). The same considerations are valid for
the Fano factors (Fig.\ref{plot2}). The straylight collected per
superpixel is expected to have a Poissonian--like statistics and
obviously increases with the binning together with the PDC light.
Therefore, its contribution, and the relative correction in the
graphs, should be independent on binning.

We want to stress once again that, differently from the case of the
flat field correction, the background can be corrected on average,
but the single image remains affected by its noise. Thus, one can
not apply any correction for the background noise to the SSNQI
scheme.

\section{quantum vs classical imaging}\label{quantum vs classical imaging}
\begin{figure}
\includegraphics[width=0.5\textwidth]{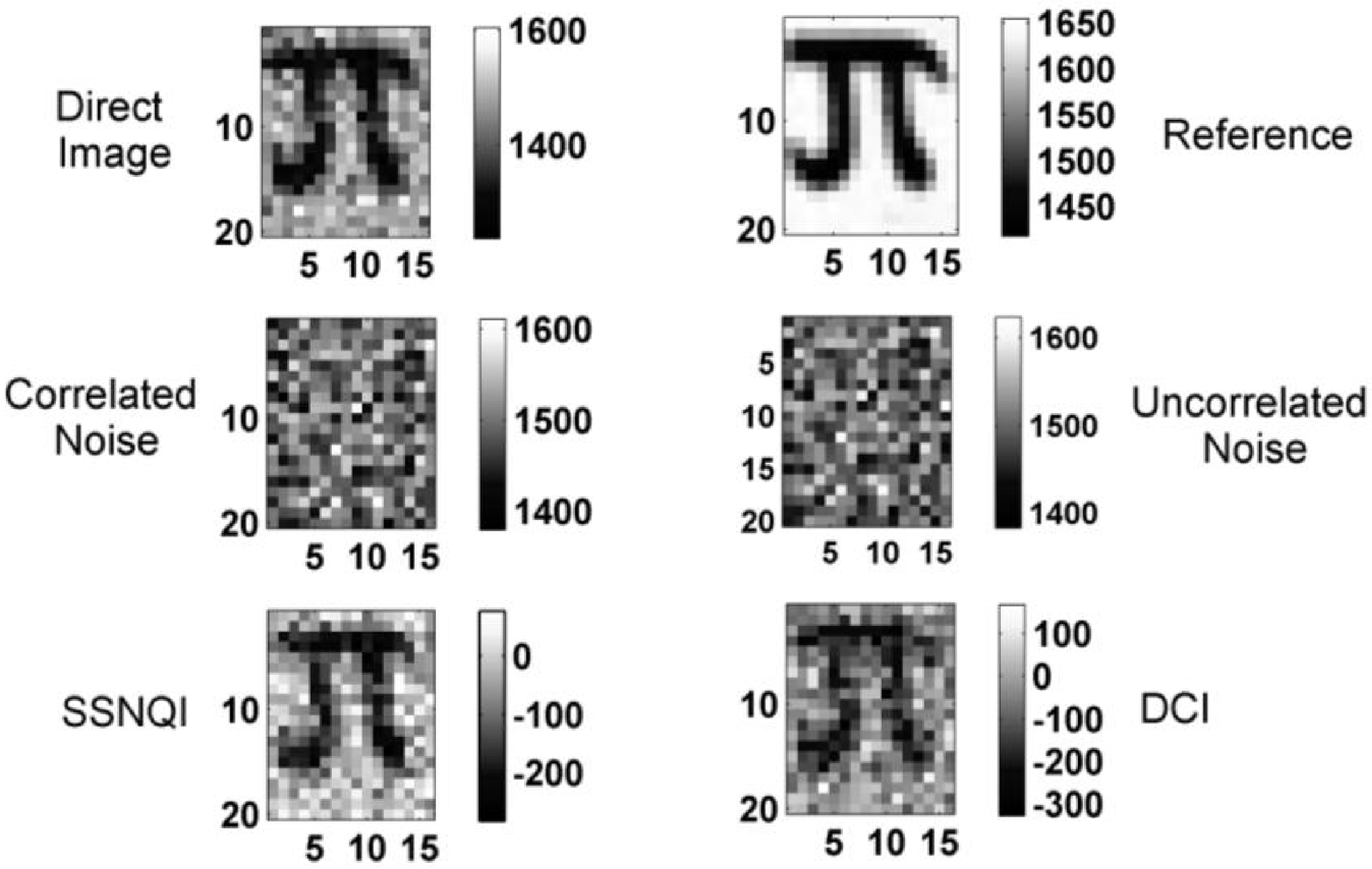}
\includegraphics[width=0.5\textwidth]{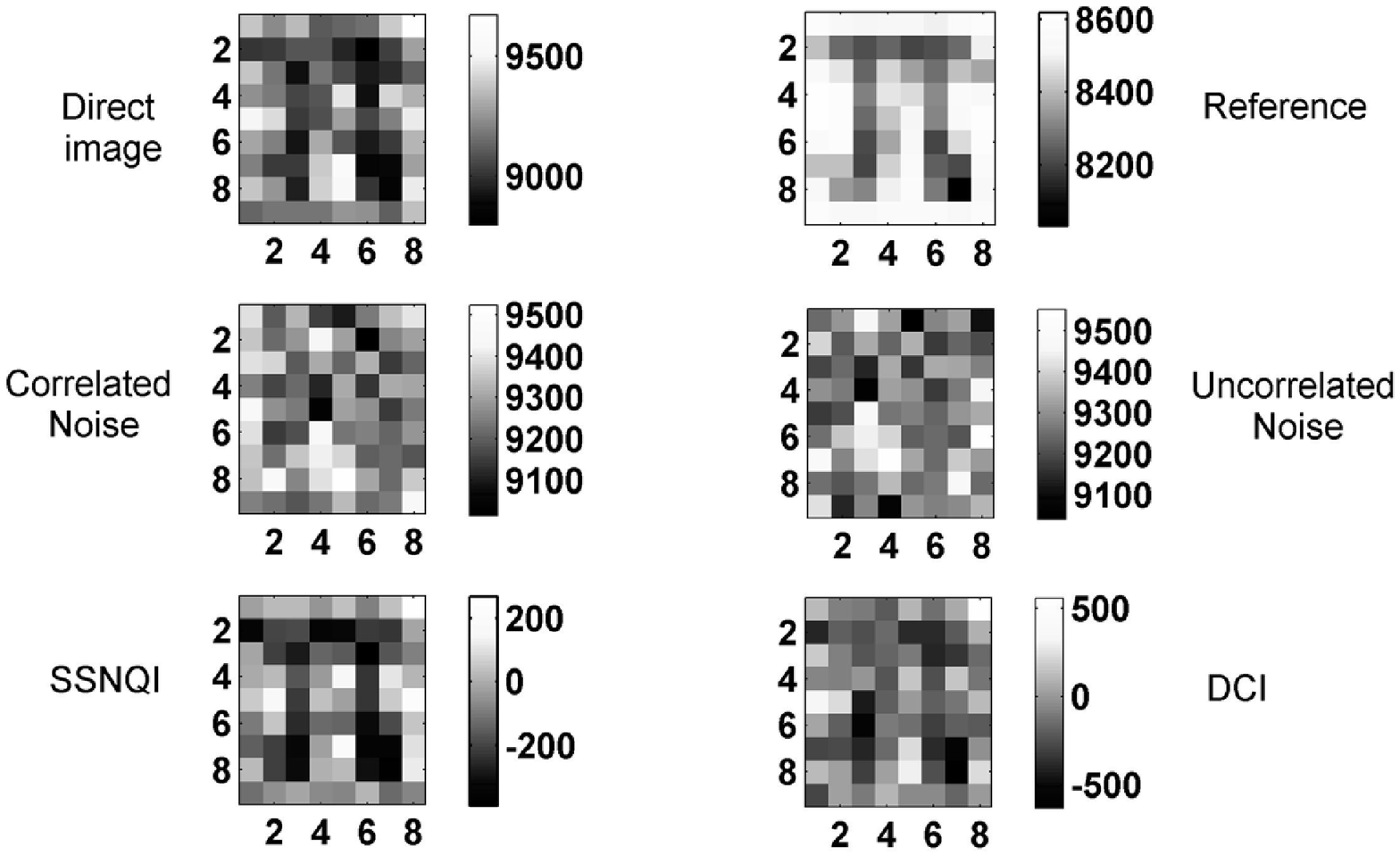}
\caption{Experimental imaging using correlated properties of PDC
quantum light, where the color-map is associated with the selected
photon number as reported on the righthand side of each image. In
(a) the pixel size is $(240\mu m)^{2}$ and  the absorption
coefficient of the object is around $\alpha=12\%$. In (b) we use
less absorbing object with $\alpha=5\%$ and pixel size of $(480\mu
m)^{2}$ (see the text for discussion).}\label{fig3}
\end{figure}

We insert a weak absorbing object in the signal beam region $R_{s}$.
It is a thin (less than 10nm) titanium deposition of 9mm$^{2}$ area
on a glass, representing a $\pi$. The SSNQI image of the object is
obtained by subtracting from the direct image registered in the
signal branch, the correlated intensity noise pattern registered in
the idler branch, as it is depicted in the left side column of
Fig.\ref{fig3}(a) for  binning 12x12 and Fig.\ref{fig3}(b) for
binning 24x24. In order to compare the results of the quantum scheme
with the corresponding DCI scheme, we simulate the last one by
subtracting from the direct image a noise pattern that is not
correlated at the quantum level with the one containing the object
(see right side of Fig.\ref{fig3} (a) and (b)). It is obtained
simply by choosing a region belonging to the idler field that is
shifted by at least one pixel with respect the quantum correlated
one. In order to simulate a coherent illumination in both branches,
the experimental spatial noise in the region $A_{s}$ and $A_{i}$
should obeys to a Poissonian statistics, i.e. it should be shot
noise limited. As we discussed above, in our working condition the
low gain and the large number of collected modes limit the spatial
excess noise of the single beams to small values (see Fano factor in
Fig.\ref{fig2} and Fig.\ref{plot2}). Furthermore, we report the
images obtained by direct classical imaging, that in the present
configuration represents the best possible classical scheme. By
observing Fig.\ref{fig3}, it is quite evident an improvement of the
SSNQI image both with respect to the DCI and direct classical one.
In the case of larger binning in particular, the object in the DCI
image is even undistinguishable ($\alpha=5\%$), whereas it is well
perceivable in the SSNQI frame. There is always a tradeoff between
the need of enlarging the detection areas for more squeezing effect,
and to preserve, at the same time the spatial resolution of the
image. In fact this is just a technical problem, that in principle
can be solved by using larger diameter of the pump laser and
crystals.

In order to quantify rigorously the improvement in the sensitivity of the SSNQI, we consider two different figures of merit.

\subsection{Correlation function}

\begin{figure}
\includegraphics[width=0.45\textwidth]{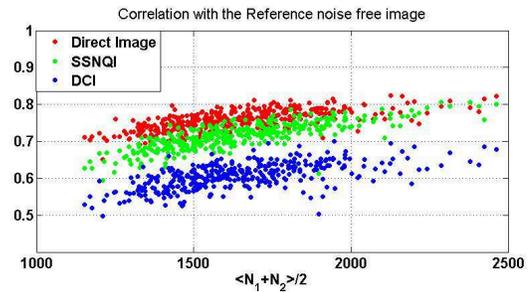}
\caption{Correlation coefficient between a reference noise free
image and the single shot frame over 400 frames. Fig. 4 present the
results for the 12x12 binning: the mean correlation coefficients are
$C_{SSQNI}=0.71$, $C_{DCI}=0.60$, $C_{direct}=0.75$ respectively for
the quantum, the differential classical, and the direct
imaging.}\label{fig4}
\end{figure}

\begin{figure}
\includegraphics[width=0.45\textwidth]{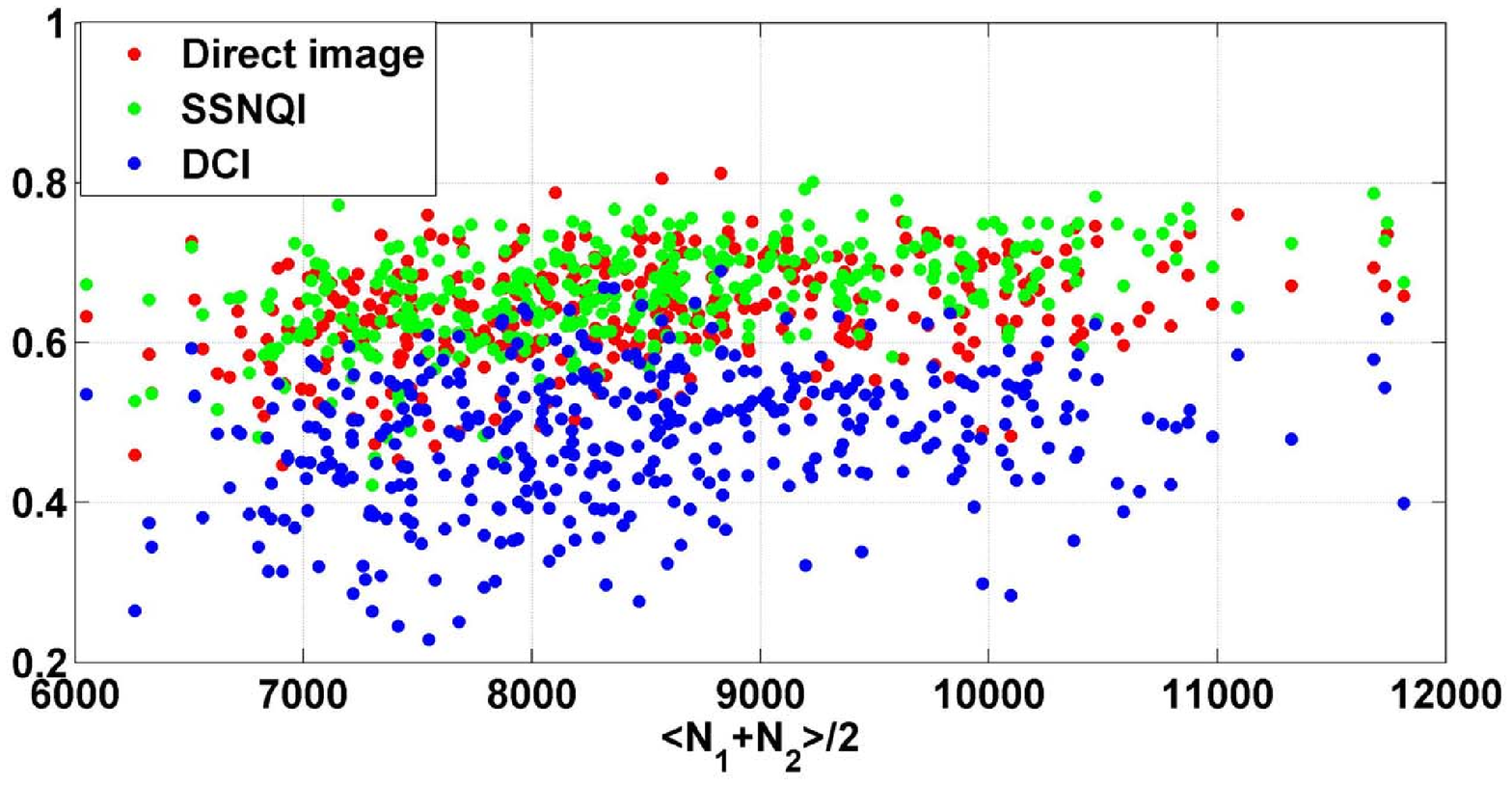}
\caption{Correlation coefficient between a reference noise free
image and the single shot frame over 400 frames. Fig. 5 present the
results for the 24x24 binning: the mean correlation coefficients are
$C_{SSQNI}=0.664$,    $C_{DCI}=0.483$, $C_{direct}=0.641$
respectively for the quantum, the differential classical, and the
direct imaging, with statistical uncertainty smaller than
$10^{-3}$}\label{fig4b}
\end{figure}

The first one consists in creating a noise free reference image of
the object, and to compare the SSNQI and the classical (DCI and
direct) images against the reference one. Operatively, the noise
free image can be prepared by averaging over a large number of
direct images, so that almost all the noise, due to multi-thermal
statistics of light as well as to the electronics of the
digitalization process, is washed away. The obtained reference
images for the two binning conditions are reported in
Fig.\ref{fig3}. The comparison of the SSNQI and the classical
schemes against the reference is quantified by the square
correlation coefficient

\begin{equation}\label{Cross2}
C_{j}=\frac{\langle \delta \mathcal{I}_{j}(\mathbf{x})\delta \mathcal{I}_{ref}(\mathbf{x})\rangle^{2}}
{\langle\delta^{2} \mathcal{I}_{j}(\mathbf{x})\rangle\langle\delta^{2}\mathcal{I}_{ref}(\mathbf{x})\rangle}
\end{equation}
where $j=SSNQI,DCI$,
$\mathcal{I}_{SSQI}(\mathbf{x})=N_{i}(-\mathbf{x})-N'_{s}(\mathbf{x})$ and
$\mathcal{I}_{DCI}(\mathbf{x})=N_{i}(-\mathbf{x+a})-N'_{s}(\mathbf{x})$.
$\mathbf{a}$ is the shift vector and the prime (') indicates the
number of photons detected in presence of the object. In principle
the image that retains more detail of the object, inspite the
deterioration due to the noise, has a larger correlation value
$C_{j}$. For instance, for the images of Fig.\ref{fig3}(b) we have
$C_{SSNQI}=0.56$, $C_{DCI}=0.41$, $C_{direct}=0.49$.

The analysis has been performed on a sample of 400 images and the
results are reported in Fig.\ref{fig4} for binning 12x12 and
Fig.\ref{fig4b} for binning 24x24. The correlation coefficient of
the SSNQI (green marks) is always larger than the one for DCI (blu
marks). In order to be completely fair one could take into account a
sample of images that presents a null excess noise on average, i.e.
a Fano factor exactly equal to 1.0 on average. In fact we did, and
the results do not differ from the behavior of the data presented in
Fig \ref{fig4} in the region of low photon number. As expected, for
the larger superpixel size, the discrimination between the classical
and the quantum imaging is more pronounced although more spread.
This is related with the degree of correlation that increases with
the detection areas, as we have seen in Sec.\ref{Experimental setup
and sub-shot noise regime}. In the case of large pixel size, the
SSNQI result is even better than the direct image obtained in the
signal branch. This is in agreement with the value of the
correlation degree that in the last case is smaller than 0.5 (see
caption of Fig. \ref{fig3}(b)). This is the evidence that the method
could becomes in the future a real tool for low noise imaging,
whenever a small illumination level is needed.

\subsection{SNR}

\begin{figure}
a)\includegraphics[width=0.5\textwidth]{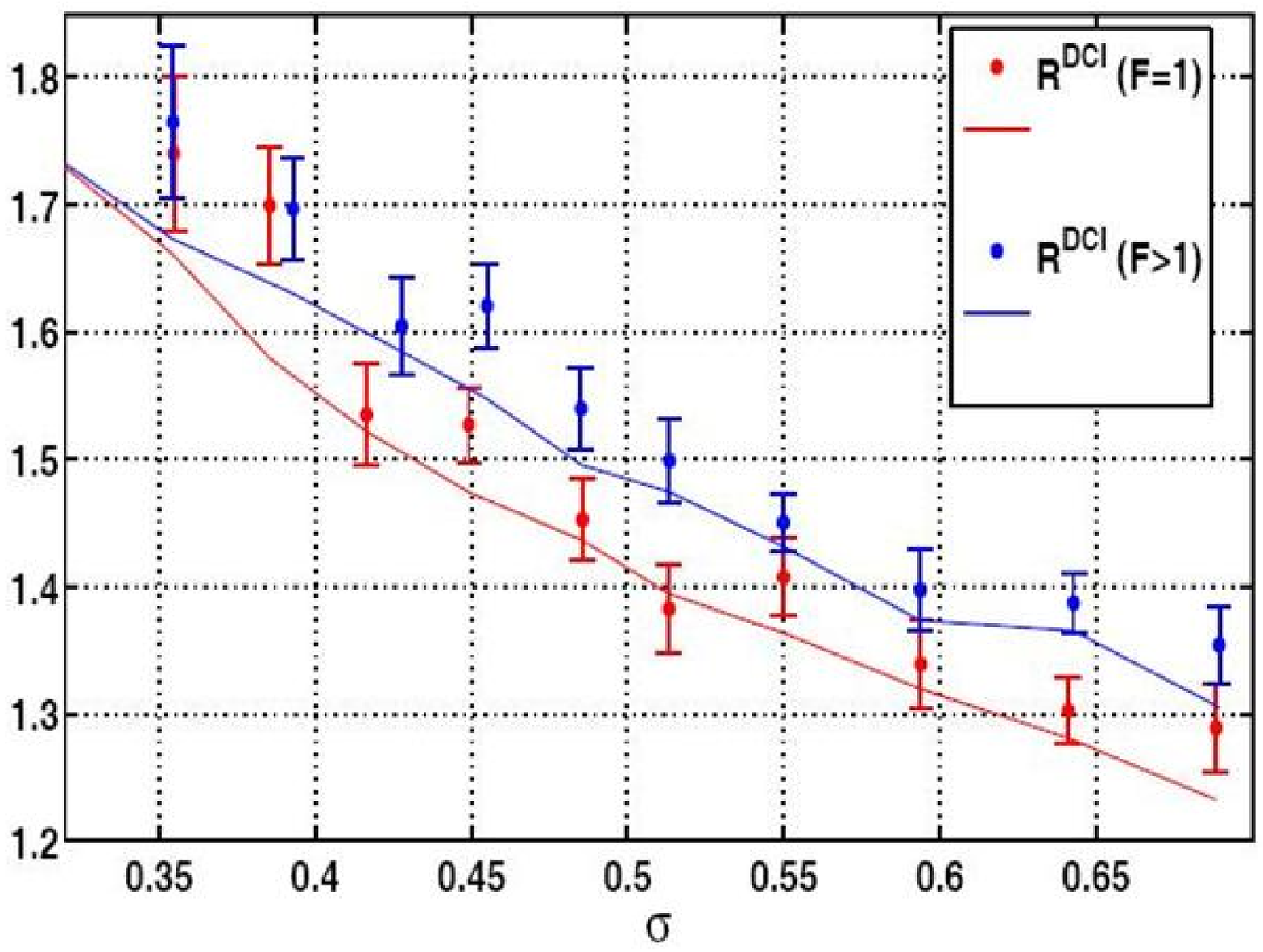}
b)\includegraphics[width=0.5\textwidth]{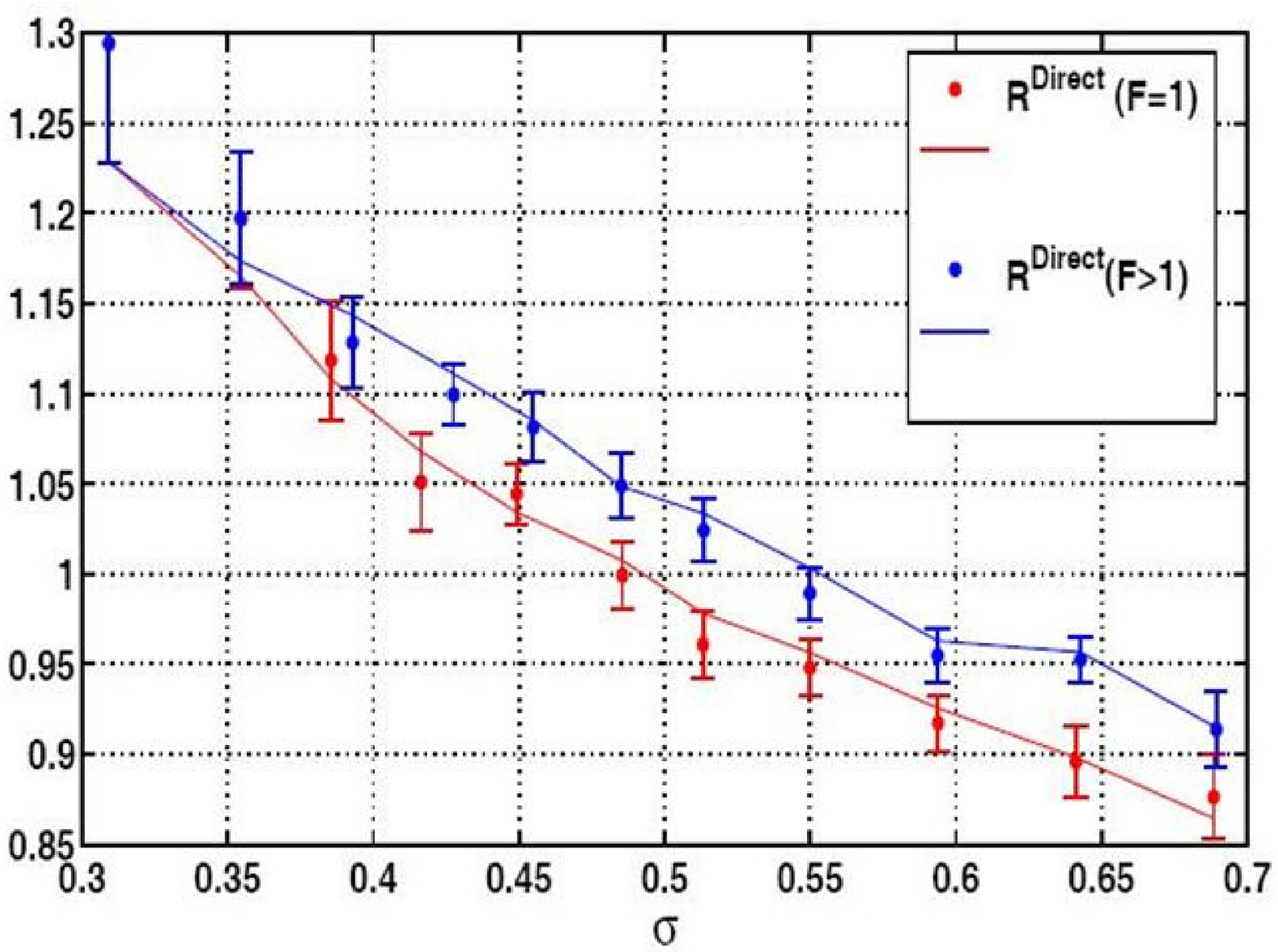} \caption{(a)
$R^{DCI}$ of the SNR in the SSNQI scheme and in the DCI scheme for
binning 24x24. The blue marks represent the result that take in to
account all the frames, without selection on the Fano factor, while
the red marks are obtained taking into account only images that have
Fano factors of $F=1\pm0.2$ (b) $R^{Direct}$ of the SNR in the SSNQI
scheme and in the Direct scheme. The lines represent the theoretical
predictions}\label{fig5}.
\end{figure}

\begin{figure}
a)\includegraphics[width=0.5\textwidth]{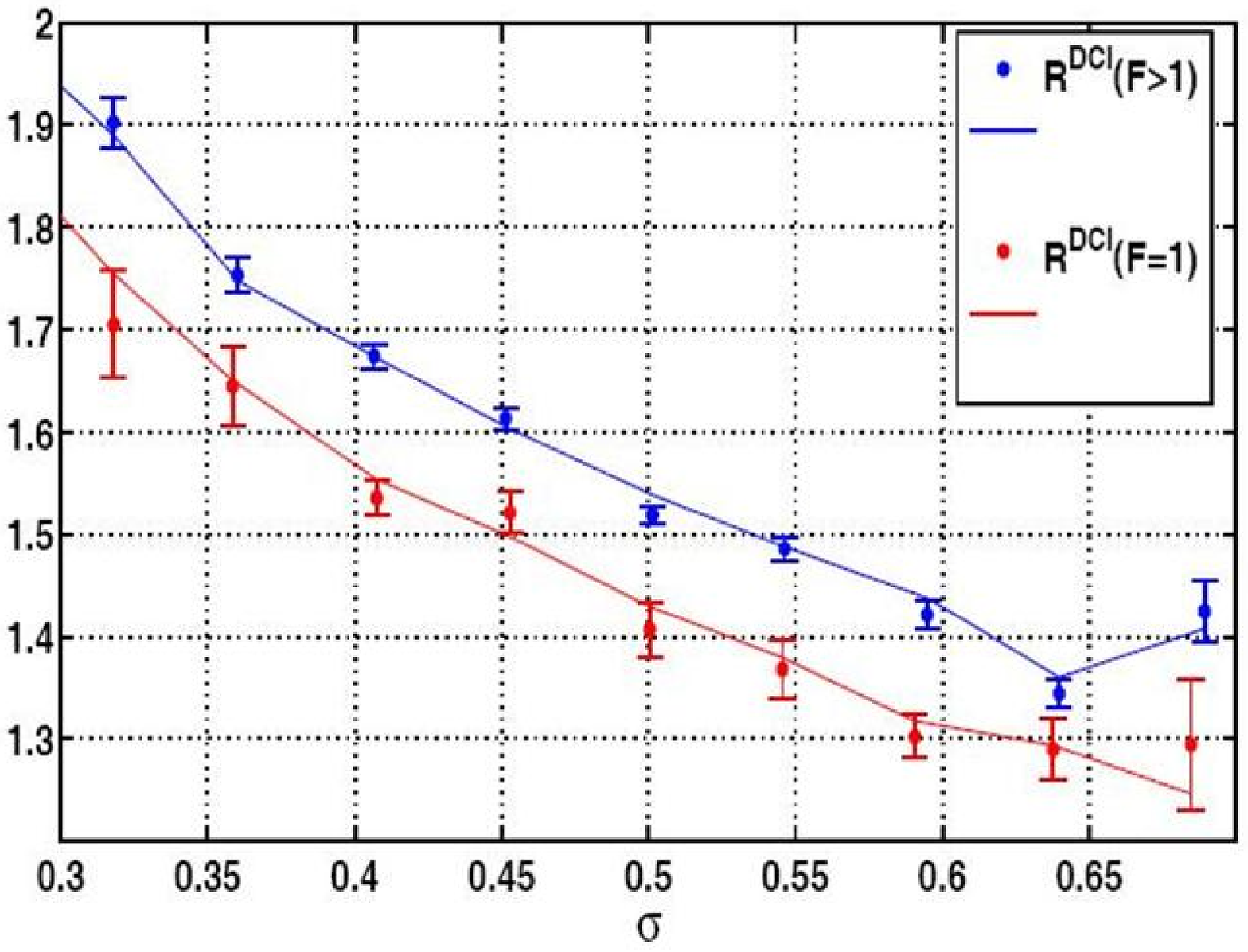}
b)\includegraphics[width=0.5\textwidth]{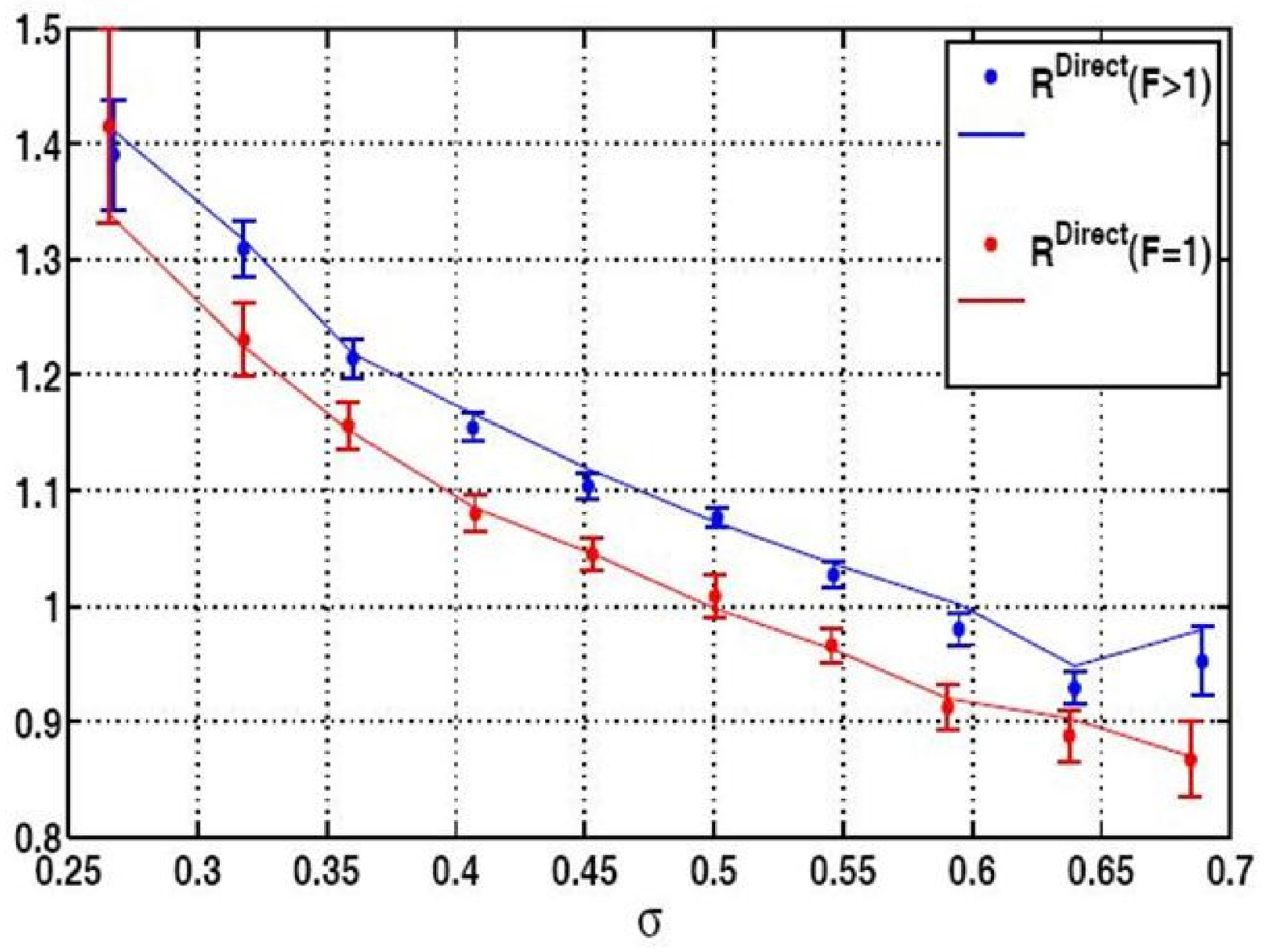} \caption{(a)
$R^{DCI}$ of the SNR in the SSNQI scheme and in the DCI scheme for
binning 32x32. The blue marks represent the result that take in to
account all the frames, without selection on the Fano factor, while
the red marks are obtained taking into account only images that have
Fano factors of $F=1\pm0.2$ (b) $R^{Direct}$ of the SNR in the SSNQI
scheme and in the Direct scheme.The lines represent the theoretical
predictions}\label{fig6}.
\end{figure}

The second figure of merit that we consider is strictly related to
the theoretical Eq.(\ref{Rdcl}). It consists in measuring the signal
to noise ratio in the classical imaging and in the quantum imaging
schemes and to compare them. For the following analysis it is
convenient to insert in the signal beam an object with a uniform
absorption $\alpha(\mathbf{x})=\alpha\sim5\%$, i.e. a square shaped
deposition that can be considered having a constant thickness for
our purposes. The prediction of Eq. (\ref{Rdcl})  are tested on  a
large number of frames (about 1200) for the case of binning 24x24
and 32x32. We group the frames in classes with respect to their
value of $\sigma$, estimated for each image as described in Sec.
\ref{Experimental setup and sub-shot noise regime}. Belonging to the
$j$- class are the images with degree of correlation falling in a
small band (bandwidth=0.1) with average $\sigma_{j}$ and average
Fano factor $F_{j}$. The estimation of the absorption spatial
distribution in the SSNQI scheme is obtained by following the Eq.
(\ref{alfa}). Experimentally, for the generic $k$-th image belonging
to the $j$-th class we perform the subtraction
$(N^{(k)}_{i}(-\mathbf{x})-N^{'(k)}_{s}(\mathbf{x}))/\langle
N^{(k)}_{i}\rangle\equiv\alpha_{k}(\mathbf{x})$ between signal and
idler correlated pattern and $\langle
N^{(k)}_{i}\rangle=\sum_{\mathbf{x}}N^{(k)}_{i}(-\mathbf{x})$. Then
the value of $\alpha(\mathbf{x})$ is obtained averaging over the
frames of the $j$-th class, i.e.
\begin{equation}\label{alfaSSQNI}
\alpha(\mathbf{x})=\frac{1}{\mathcal{N}_{j}}\sum^{\mathcal{N}_{j}}_{k=1}\alpha_{k}(\mathbf{x})\equiv\overline{\alpha_{k}(\mathbf{x})}
\end{equation}
where $\mathcal{N}_{j}$ is the number of images in the $j$-th class.

The differential classical scheme with coherent beam can be
simulated in our set-up by evaluating the absorption as in the
previous Eq. (\ref{alfaSSQNI}) where the region of the idler is
slightly shifted from the quantum correlated one
$N_{i}(-\mathbf{x}+\mathbf{a})$  (typically we take
$\mathbf{a}=(1,0))$.

For each class, the SNR for the quantum and for the classical
schemes can be estimated punctually as $SNR_{j}
(\mathbf{x})=\overline{\alpha_{k}(\mathbf{x})}/[\overline{\delta^{2}\alpha_{k}(\mathbf{x})}]^{1/2}$,
where the overline indicates the average over the frames of the
class $j$. At the same time we evaluate the factor
$R_{j}^{DCI}(\mathbf{x})=SNR_{j}^{SSNQI} /SNR_{j}^{DCI}$. Then, we
perform also a spatial averages over $\mathbf{x}$, to obtain a
representative value of $R_{j}^{DCI}$ for each class. The results
are shown in Fig.\ref{fig5}(a) and Fig.\ref{fig6}(a) against the
correlation degree $\sigma_{j}$ for the binning 24x24 and 32x32
respectively: the SSNQI has, as expected, an advantage that
increases with the level of correlation. Selecting only images with
Fano factors very close to 1 for simulating a classical coherent
illumination (red data-set), the result are just slightly worse as
one without Fano selection (blu data-set). However, for the best
achieved values of the correlation degree in the quantum regime, we
obtain an improvement of the SNR larger than than 70\% compared with
the classical imaging scheme. The results are in agreement with the
theoretical predictions obtained by the generalization of Eq.
(\ref{Rdcl}) with the substitutions $E^{class}_n\rightarrow F_{j}-1$
and $\sigma\rightarrow \sigma_{j}$.

In Fig. \ref{fig5}(b) and Fig.\ref{fig6}(b) we report the Ratio $R$
of the SNR in the SSNQI scheme and in the direct classical scheme.
When Fano values are selected to be $\sim1$ (red data-set), allowing
a proper simulation of a shot noise limited classical imaging, the
results are in perfect agreement with what expected from theory. In
particular for small $\alpha$, we have $R={1 \over \sqrt{2
\sigma}}$, i.e. the SSQI overcomes the direct classical imaging as
soon as $\sigma < 0.5$.

\section{Conclusion}\label{Conclusion}
We have presented the first experimental quantum imaging protocol
aimed to improve the sensitivity of imaging techniques beyond the
shot-noise-limit, that represents the threshold between classical
and quantum world. We reached a considerably high level of
multi-mode spatial correlations using the properties of twin beams
generated by the parametric-down-conversion process, and we used it
for reducing the noise in the differential imaging scheme. Our
results represent the proof of principle of this new quantum
technology. Anyway, some  technical limitation toward practical
application of the methods are still present in our setup. Firstly,
total transmission of the optical channel should be improved in
order to reach even higher quantum correlations. Furthermore, there
is always a tradeoff between the need of enlarging the detection
areas for more squeezing effect, and at the same time to preserve
the spatial resolution of the image. In fact this is just a
technical problem, that in principle can be solved by reducing the
coherence area of the single spatial mode, in particular using
larger diameter of the pump laser and crystals. A work for solving
these problems in view of practical applications in now going on
\cite{iv}.

\section{Acknowledgements} This work has been supported by PRIN 2007FYETBY (CCQOTS).
Thanks are due to E.Monticone and C. Portesi for the thin film
deposition representing the weak absorbing object, to P. Cadinu for
help in the data analysis and to A. Gatti, E. Brambilla, L. Caspani
and L. Lugiato for theoretical support.



\begin{thebibliography}{99}
\bibitem{Commento} In a single pixel $k$ are collected $\mathcal{M}$ modes of the
radiation. Since the mean value of the number of photon per modes
$\langle \hat{\mathfrak{n}}_m\rangle$ is considered as constant, the
average number of photon per pixel is $\langle \hat{n} \rangle
=\sum_{m=1}^{\mathcal{M}}\langle \hat{\mathfrak{n}}_m
\rangle=\mathcal{M}\langle \hat{\mathfrak{n}} \rangle$. Assuming
that the modes are statistically independent and that the number of
photons per mode in each branch obeys to a thermal statistics
\cite{mandel}, its variance is: $\langle \delta^2 \hat{n} \rangle
=\sum_{m=1}^{\mathcal{M}}\langle \delta^2 \hat{\mathfrak{n}}_m
\rangle=\mathcal{M}\langle \delta^2 \hat{\mathfrak{n}} \rangle =
\mathcal{M}\langle  \hat{\mathfrak{n}} \rangle(1+\langle
\hat{\mathfrak{n}}\rangle)=\langle \hat{n}
\rangle\left(1+\frac{\langle \hat{n} \rangle}{\mathcal{M}}\right)$.
\end{thebibliography}
\end{document}